\documentstyle[twoside,fleqn,espcrc2]{article}

\newcommand{\AmS}{{\protect\the\textfont2
  A\kern-.1667em\lower.5ex\hbox{M}\kern-.125emS}}

\title{Universal Amplitude Ratios in the 3D Ising Model
       }

\author{M. Caselle\address{Istituto Nazionale di Fisica Nucleare, 
 Sezione di Torino, 
via P.Giuria 1, I-10125 Torino, Italy }
	and M. Hasenbusch\address{ 
Humboldt Universit\"at zu Berlin,
	Institut f\"ur Physik, Invalidenstr. 110, D-10115 Berlin, Germany}
\thanks{Talk presented by M.Hasenbusch}
}
\begin{document}

\begin{abstract}
We present a high precision Monte Carlo study of various
universal amplitude
ratios of  the three dimensional Ising spin model.
Using state of the art simulation
techniques we studied the model close to criticality in both phases.
Great care was taken to control systematic errors due to finite
size effects
and correction to scaling terms. We obtain  $C_+/C_-=4.75(3)$,
$f_{+,2nd}/f_{-,2nd}=1.95(2)$ and $u^*=14.3(1)$.
Our results are compatible with those obtained by field theoretic methods
applied to the $\phi^4$ theory  and high and low temperature
series expansions of the Ising model.
\end{abstract}

\maketitle

\newcommand{\eq}{\begin{equation}}
\newcommand{\en}{\end{equation}}
\newcommand{\eqa}{\begin{eqnarray}}
\newcommand{\ena}{\end{eqnarray}}
\newcommand{\spz}{\hspace{0.7cm}}
\newcommand{\lbl}{\label}
\newcommand{\lhi}{\hat\lambda_{i}}
\newcommand{\br}{\langle}
\newcommand{\kt}{\rangle}
\newcommand{\um}{\frac12}
\newcommand{\th}[1]{\vartheta_{#1}(0,\tau)}


\newcommand{\NP}[1]{Nucl.\ Phys.\ {\bf #1}}
\newcommand{\PL}[1]{Phys.\ Lett.\ {\bf #1}}
\newcommand{\NC}[1]{Nuovo Cim.\ {\bf #1}}
\newcommand{\CMP}[1]{Comm.\ Math.\ Phys.\ {\bf #1}}
\newcommand{\PR}[1]{Phys.\ Rev.\ {\bf #1}}
\newcommand{\PRL}[1]{Phys.\ Rev.\ Lett.\ {\bf #1}}
\newcommand{\MPL}[1]{Mod.\ Phys.\ Lett.\ {\bf #1}}
\newcommand{\IJMP}[1]{Int.\ J.\ Mod.\ Phys.\ {\bf #1}}
\newcommand{\JETP}[1]{Sov.\ Phys.\ JETP {\bf #1}}
\newcommand{\TMP}[1]{Teor.\ Mat.\ Fiz.\ {\bf #1}}

\section{Introduction}
 In the neighbourhood of a second order phase transition
 various quantities display
 a singular behaviour. In this limit  microscopic features 
 become irrelevant and models which 
 differ at the microscopic level may share the same singular behaviour.
 This is the basis of universality.
 Different models belonging to the same universality class share the same
 critical indices. 
 However universality has much stronger implications
 and it is possible to show that for a given universality class
 the values of particular critical-point amplitude 
 combinations are unique~\cite{ahp}. 
In the following we shall be interested in the universality class of
the 3 dimensional Ising model which has several interesting experimental
realizations, ranging from the binary mixtures to the liquid vapour
transitions.
The action of the Ising model is given by
\eq
S = - \beta \sum_{<n,m>} s_n s_m \; ,
\label{Sspin}
\en
where
  $s_n \in \{-1,1\}$ is the field variable and
$<n,m>$ is a pair of nearest neighbour sites on the lattice. 
$\beta\equiv \frac{1}{kT}$ is the coupling.
The reduced temperature $t$ can be written as
$t=\frac{\beta_c-\beta}{\beta}$
where $\beta_c$ is the critical coupling.
In the following we  consider lattices of
volume $L^3$ and periodic boundary conditions.

We simulated the model in the low as well as in the high temperature phase.
The simulations in the high temperature phase have been performed using
Wolff's single cluster algorithm.
Here the improved estimators give a great boost to the 
accuracy of the results. 
However in the low temperature phase, due to the finite magnetisation,
the improved
estimators of the cluster-algorithm are of little help. Hence we simulated here 
with a multi-spin coded demon algorithm. 

We measured the magnetisation, the magnetic susceptibility $\chi$ 
and the exponential
$\xi$ as well as the second moment correlation length $\xi_{2nd}$.
In order to obtain a finite result for the magnetisation in the low 
temperature phase on finite lattices
at vanishing external field we measured 
\begin{equation}
m = \frac1V <|\sum_n s_n|>   \;\;\; .
\end{equation}
The magnetisation defined in this way was also used for the subtraction
in connected correlation functions in the low temperature phase.

In the neighbourhood of the critical point one expects a singular behaviour
of the form $m \sim B \; t^{\beta}$, $ \chi \sim C_{\pm} |t|^{-\gamma}$,
$\xi  \sim f_{\pm} |t|^{-\nu}$ and 
$\xi_{2nd}  \sim f_{\pm,2nd} |t|^{-\nu}$,
where $+$ indicates the high temperature phase $t>0$ and $-$ the  low
temperature phase $t<0$.

Ratios and more general combinations of singular quantities where 
the singularities cancel are expected to be universal. In our study we
computed $C_+/C_-$,
$\; f_{+,2nd}/f_{-,2nd}$, $\; 3 \; C_-/f_{-,2nd}^3 B^2$ 
and $C_+/f_{+,2nd}^3 B^2$. The finiteness of the last two quantities
relies on the scaling (hyper-scaling) relations $\alpha + 2 \beta + \gamma
=2 $ and $d \nu = 2- \alpha$.

\section{numerical results}
First we simulated the model in the low temperature phase at 
$\beta=0.2391$,
$0.23142$, $0.2275$, $0.2260$,$0.2240$ and $0.22311$ on lattices 
of size $L=30$ up to  $L=120$. From finite size tests at the largest
$\beta$ we concluded that $L > 20 \xi$ is needed to be save of finite
size effects. The total number of measurements was about $3\times 10^6$ 
for all $\beta$'s. We carefully compared our results with IDA's of 
low temperature series expansions and Monte Carlo results
given in the literature. In general we found good agreement. For details
see ref. \cite{us}.

Next we simulated the model in the high temperature phase at
$\beta=0.20421$, $0.21189$, $0.21581$, $0.21731$, $0.21931$ and $0.22020$.
In the high temperature phase only $L > 6 \xi$ is required to be 
sufficiently close to the thermodynamic limit.
The $\beta$-values 
were chosen such that they pair up with a $\beta$ in the 
low temperature phase such that $\beta_{low} + \beta_{high} = 2\beta_c$.

This choice of $\beta$-values allows to compute for example approximations 
of $C_+/C_-$ by
$\Gamma_{\chi}(t) = \frac{\chi(t)}{\chi(-t)} $
with $t>0$.  In this way we need not to extract the $C_{\pm}$ themselves and
the critical exponent $\gamma$ does not enter into the calculation.
The critical limit is then obtained from a fit to the ansatz
\begin{equation}
\label{fit2}
\Gamma_{\chi}(t) \; = \; C_+/C_- \;+\; a_0 
\;t^{\theta}
 \; +\; a_1 \; t \;\;\; .
\end{equation}
where we  have included the leading corrections to scaling. The best 
know numerical estimate of the correction exponent is
$\theta=0.51(3)$.
The results for the other amplitude ratios and combinations are extracted 
analogously. 
In table 1 we have summarised the approximations of $C_+/C_-$,
$\; f_{+,2nd}/f_{-,2nd}$ and $u^*=3 C_-/f_{-,2nd}^3 B^2$ 
for finite $t$. The last line gives our result for the extrapolation
to the critical limit. 
In addition we have $C_+/f_{+,2nd}^3 B^2=3.05(5)$
The largest reduced temperature $t$ is excluded from the fit in all cases.
The errors quoted for the critical limit include 
errors induced by the uncertainty of $\beta_c$ and $\theta$. The 
$\chi^2/d.o.f.$ was of order $1$  for all quantities.

\begin{table}[h]
\label{datafit}
\caption{\sl The various ratios as  functions of the reduced temperature $t$
The last line gives the extrapolation to the critical limit}
\vskip 0.2cm
\begin{tabular}{cccc}
\hline
$t$  &  $\Gamma_{\chi}$ &  $\Gamma_{\xi}$ & $u$  \\
\hline
0.0787 &  6.044(5) &  1.902(2) &  15.00(6) \\
0.0441 &  5.546(5) &  1.920(3) &  14.75(5) \\
0.0264 &  5.283(3) &  1.932(3) &  14.66(6) \\
0.0196 &  5.182(5) &  1.939(3)&  14.64(5) \\
0.0106 &  5.027(6) &  1.942(3) &  14.47(6)\\
0.0066 &  4.947(6) &  1.948(3) &  14.51(7) \\
\hline   
0      &   4.75(3)      & 1.95(2) &  14.3(1) \\ 
\hline
\end{tabular}
\end{table}

\begin{table}[h]
\label{literature2}
\caption{\sl Results for the amplitude ratios reported 
 in the literature. They are obtained by $\epsilon$-expansion~[2],
 perturbation theory in 3 dimensions~[2,3],
 high and low temperature series expansions of the Ising model~[4] 
and  Monte Carlo simulations~[5].
}
\vskip 0.2cm
\begin{tabular}{ccccc}
\hline
ref.&  $\frac{C_+}{C_-}$ & $\frac{f_{+,2nd}}{f_{-,2nd}}$
 & $u^*$  \\
\hline
\cite{gz} &  4.70(10) & & \\  
\cite{gz} & 4.82(10) & & & \\
\cite{munster2}& 4.72(17) & 2.013(28) & 
14.4(2)  \\
\cite{lf} &  4.95(15) & 1.96(1) 
&14.8(1.0) \\
\cite{kielxi}& 5.18(33) &2.06(1) &
17.1(1.9) \\
\hline
\end{tabular}
\end{table}

\section{Comparison with theoretical results and experimental data}
In table 2 we have summarised the most recent results for amplitude 
ratios and combinations for the 3D Ising universality class obtained from
$\epsilon$-expansion \cite{gz}, 
perturbation theory in 3 dimensions \cite{gz,munster2}, high and low
temperature series expansions of the Ising model \cite{lf}
and a Monte Carlo 
study of the Ising model \cite{kielxi}.  The results for $C_+/C_-$ of refs.
\cite{gz,munster2,lf,kielxi}
are in good agreement with our estimate. Our result is however
more accurate.  For $f_{2nd,+}/f_{2nd,-}$ our result is compatible
with that of \cite{lf},
the result from perturbation theory in 3 dimensions \cite{munster2} is 
larger then our estimate, however still within a 2$\sigma$ deviation.
There is however a clear discrapency with the Monte Carlo result of
ref. \cite{kielxi}. Their value is much larger then our one.
For $u^*$ and $\frac{C_+}{f^3_{+,2nd}B^2}$ our result is consistent
with the other methods. In both cases we could reduce the errors 
considerably.

The experimental data reported in tab.3 refer to the three most
important experimental realizations of the Ising universality class, 
namely 
binary mixtures (bm),  liquid-vapour transitions (lvt) and uniaxial
 anti-ferromagnetic systems (af). 
It is important to notice that these realizations are not on the
same ground. Antiferromagnetic systems are particularly apt to measure the
$C_+/C_-$ and $f_{+,2nd}/f_{-,2nd}$ ratios, while for
 the liquid vapour transitions 
the $\Gamma_c\equiv R_c/R_\xi^3$ combination is more easily accessible.
Finally, in the case of binary mixture all the three ratios can be rather
easily evaluated. 
The common attitude is to assume
 that the above systematic
errors are randomly distributed and to take
 the weighted mean of the various experimental results. 
The numbers are taken from ref.
\cite{ahp}, where more details on the experiments and the averaging 
procedure can be found.

\begin{table}[h]
\label{literature3}
\caption{\sl Experimental estimates for some amplitude ratios.
(bm) refers to binary mixtures, (lvt) to  liquid-vapour transitions and
(af) to uniaxial anti-ferromagnetic systems. (all of them) gives an 
weighted average.
}
\vskip 0.2cm
\begin{tabular}{cccc}
\hline
exp. setup  & $\frac{C_+}{C_-}$ & $\frac{f_{+,2nd}}{f_{-,2nd}}$
  & $\frac{C_+}{f^3_{+,2nd}B^2}$  \\
\hline
 (bm)  & 4.4(4) & 1.93(7) & 3.01(50) \\
 (lvt) & 4.9(2) & &  2.83(31)\\
 (af)  & 5.1(6) & 1.92(15)  &   \\
 (all of them) &4.86(46) & 1.93(12) & 2.93(41)  \\
\hline
\end{tabular}
\end{table}

In the case of the $f_{+,2nd}/f_{-,2nd}$ ratio (for which, as we have seen, 
 some of the present theoretical or Monte Carlo estimates disagree)
 we have listed, for a more detailed comparison, all the available 
experimental data in tab.17. In this table we denote with 
``N-H'' the nitrobenzene~--~n-hexane binary mixture, and with
 ``I-W'' the one obtained by mixing isobutyric acid and water.
 A much more detailed account of the various 
experimental estimates can be found in~\cite{ahp}.

\begin{table}[h]
\label{literature4}
\caption{\sl Experimental estimates for the $\frac{f_{+,2nd}}{f_{-,2nd}}$
ratio}
\vskip 0.2cm
\begin{tabular}{cccc}
\hline
Ref. & year & exp. setup &  $\frac{f_{+,2nd}}{f_{-,2nd}}$ \\
\hline
\cite{hsg} & 1972  & (af),${\rm FeF_2}$ & 2.06(20) \\
\cite{cc} & 1980  & (af),${\rm CoF_2}$ & 1.93(10) \\
\cite{zbb} & 1983  & (bm),N-H & 1.9(2) \\
\cite{htkk} & 1986  & (bm),I-W & 2.0(4) \\
\hline
\end{tabular}
\end{table}

A detailed account of our work can be found in ref \cite{us}.
%


\begin{thebibliography}{99}

\bibitem{ahp} For a comprehensive review, see for instance: 
A. Aharony, P.C. Hohenberg and V.Privman,
{\em Universal critical point amplitude relations} in ``Phase transitions and
Critical phenomena'' vol.14, C.Domb and J.L. Lebowitz eds. (Academic Press 1991)


\bibitem{gz} R.Guida and J.Zinn-Justin \NP{B489} (1997) 626.
\bibitem{munster2}  C. Gutsfeld, J. K\"uster and G. M\"unster, 
                     \NP{B479} (1996) 654.
%
%
\bibitem{lf}{A.J. Liu and M.E. Fisher, Physica {\bf A 156} (1989) 35.}


\bibitem{kielxi}  C. Ruge, P. Zhu and F. Wagner, Physica {\bf A 209}
 (1994) 431.







{M.E. Fisher and H. Au-Yang, J. Phys. {\bf A 12} (1979) 1677.}


%













\bibitem{hsg} M.T.Hutchings, M.P.Schulhof and \\ H.J.Guggenheim
   \PR{B5} (1972) 154.
\bibitem{cc} R.A.Cowley and K.Carniero J. Phys. {\bf C13} (1980) 3281.
\bibitem{zbb} G. Zalczer, A.Bourgou and D. Beysens   
\PR{A28} (1983) 440.
\bibitem{htkk} K. Hamano, S.Teshigawara, T.Koyama and N.Kuwahara,
\PR{A33} (1986) 485.

\bibitem{us} M. Caselle and M. Hasenbusch,
J.Phys.{\bf A 30} (1997) 4963.

\end{thebibliography}
\end{document}